\documentclass[pra,letterpaper,10pt,twocolumn]{revtex4-1}
\usepackage{graphicx}
\usepackage[final,dvips]{epsfig}
\usepackage{color}
\usepackage{dcolumn}  
\usepackage{bm}       
\usepackage{array}    
\usepackage{amssymb,amsmath}
\usepackage{wasysym}
\usepackage{appendix}
\usepackage{booktabs}
\usepackage[utf8x]{inputenc}
\usepackage[colorlinks,citecolor=blue]{hyperref}
\emergencystretch=\maxdimen
\newcommand{\be}{\begin{equation}}
\newcommand{\ee}{\end{equation}}
\newcommand{\ba}{\begin{eqnarray}}
\newcommand{\ea}{\end{eqnarray}}

\newcommand{\upa}{\uparrow}
\newcommand{\dna}{\downarrow}

\newcommand{\COMMENTED}[1]{}

\newcommand{\ket}[1]{|#1\rangle}

\newcommand{\ob}[1]{{\langle #1\rangle}}

\newcommand{\bfi}{\mathbf{i}}
\newcommand{\bfj}{\mathbf{j}}

\newcommand{\bfr}{\mathbf{r}}

\newcommand{\bfS}{\mathbf{S}}

\newcommand{\hH}{{\hat{H}}}

\newcommand{\hS}{{\hat{S}}}
\newcommand{\hT}{{\hat{T}}}

\newcommand{\hV}{{\hat{V}}}

\newcommand{\hb}{{\hat{b}}}

\newcommand{\hn}{{\hat{n}}}

\usepackage{natbib}

\begin{document}

\title{Competing exotic quantum phases of spin-1/2 ultra-cold lattice
bosons with extended spin interactions}

\author{Chia-Chen Chang${}^1$}
\author{Val\'ery G. Rousseau${}^2$}
\author{Richard T. Scalettar${}^1$}
\author{George G. Batrouni${}^{3,4,5,6}$}
\affiliation{
${}^1$Department of Physics, University of California Davis, 95616 CA  USA\\
${}^2$Department of Physics, Louisiana State University, Baton Rouge, 70803 LA, USA\\
${}^3$INLN, Universit\'e de Nice -- Sophia Antipolis, CNRS; 1361 route
des Lucioles, 06560 Valbonne, France\\ 
${}^4$Institut Universitaire de France, 103, Boulevard Saint-Michel,
75005 Paris, France\\ 
${}^5$Centre for Quantum Technologies, National University of
  Singapore, 3 Science Drive 2, Singapore 117543\\
${}^6$Merlion MajuLab, CNRS-UNS-NUS-NTU International Joint
  Research Unit UMI 3654, Singapore
}

\begin{abstract}
  Advances in pure optical trapping techniques now allow the 
  creation of degenerate Bose gases with internal degrees of
  freedom.  Systems such as ${}^{87}$Rb, $^{39}$K or ${}^{23}$Na in
  the $F=1$ hyperfine state offer an ideal platform for studying the
  interplay of superfluidity and quantum magnetism.
  Motivated by the experimental developments, we study ground state
  phases of a two-component Bose gas loaded on an optical lattice. The
  system is described effectively by the Bose-Hubbard Hamiltonian with
  onsite and near neighbor spin-spin interactions.  An important
  feature of our investigation is the inclusion of interconversion
  (spin flip) terms between the two species,
  which has been observed in optical lattice experiments.
  Using mean-field theory
  and quantum Monte Carlo simulations, we map out the phase diagram of
  the system. A rich variety of phases is identified, including
  antiferromagnetic (AF) Mott insulators, ferromagnetic and AF
  superfluids.
\end{abstract}

\maketitle 

\section{Introduction}

The question of the interplay of superfluidity and internal bosonic
degrees of freedom dates back many decades.  At a purely conceptual
level, the most straightforward issue is whether the two internal
components move in, or out of phase.  A generalization of Bogoliubov's
treatment to multiple species addressed this question and demonstrated
that a neutral mixture of two species of charged bosons supports
plasma-type excitations with oscillating charge density and also
free-particle oscillations associated with mass density
oscillations\cite{Bassichis64}.  An extension of this work to finite
temperatures considered dilute mixtures of (unstable) $^6$He in $^4$He
\cite{Colson78}.  Coupling between bosonic species was also shown to
imply that superfluid motion of one component would result in a ``drag
effect'' in which the second component is also set in
motion\cite{Nepomnyashchii64}.

Although different theoretical and experimental motivations were
presented for studying multi-component bosons, in this early work, the
physical system which was probably considered in the most detail was
spin polarized hydrogen\cite{Hecht1959,Etters1975,Stwalley1976}, where
a large external magnetic field prevents recombination into molecules,
and the smallness of the kinetic energy relative to the binding energy
permits treatment with boson statistics.  The key
observation\cite{Siggia1980} was that these bosons reside in two
low-lying hyperfine states, thus allowing for possible additional
symmetry breaking associated with their relative occupation. The
populations of the states were measured with electron spin
resonance\cite{Yperen1983}, and the nature of excitations of
the ``spin'' degrees of freedom was shown to range from
phonon-like, to free-particle-like with energy gaps, to resembling
spin waves\cite{Berlinsky1977}. For a review, see
Ref.~[\onlinecite{Greytak1984}].

Lattice models were also investigated.  A mean field
treatment\cite{Fazekas1983} of the hard-core limit of two component
bosons focussed on the effect of ``antiferromagnetic'' interactions,
i.e. a repulsion $V$ between the different bosonic species on
near-neighbor sites of a bipartite structure.  Besides promoting an
insulating phase where bosonic species alternate in a regular pattern,
this interaction was found also to disrupt the ``symmetrical
condensate''\footnote{The authors of Ref.~\onlinecite{Fazekas1983}
  employ the term ``symmetric'' to emphasize the presence of axial
  symmetry about the ``magnetic field'' direction (in boson language,
  the direction of the term which provides a different energy to the
  two species).}
in which only a single superfluid species occurs, and
allow for a superfluid phase in which both species are present. It was
shown that, as a function of temperature $T$, two successive second
order transitions can occur. For sufficiently large $V$, as $T$ is
lowered, the bosons first form a symmetric condensate and then, at a
distinct, lower temperature, the asymmetrical condensate appears.  We
will show here that the soft-core lattice problem exhibits certain
similarities with these hard core phase diagrams.

Beginning in the late 1990's, the properties of multicomponent boson
systems became of renewed interest due to applications to ultracold
quantum gases. 
Thanks to the all-optical trapping technique\cite{stamperkurn98},
hyperfine states of $^{87}$Rb, $^{39}$K or $^{23}$Na in optical traps 
could now be used to realize interesting magnetic states\cite{stamperkurn98,ho98,Ohmi:1998}.
In an optical lattice, it has been observed that atoms confined 
on the same lattice site exhibits collisions that could change their spin 
states\cite{Widera:2005}. Such a system can be effectively described
by a multi-component Bose-Hubbard model with appropriate values of 
the intra- and inter-component interactions, and spin-conversion 
matrix elements\cite{Krutitsky:2004,Krutitsky:2005}. 
Due to the competition between spin species, the multi-component
Bose-Hubbard model is expected to host novel 
phases\cite{Lewenstein:2012,StamperKurn:2013,Krutitsky:2015} 
that are absent in the one-component Bose-Hubbard model\cite{Fisher:1989}.

Motivated by these experimental and theoretical developments for
spinor bosons in optical lattices,
we will study two component (spin-$1/2$) bosons on a
two-dimensional lattice.  We consider a very general Hamiltonian which
includes not only on-site repulsion, but also near-neighbor
interactions and interconversion between the species through a
spin-spin coupling.  We begin with a mean field theory (MFT) treatment
which reveals a rich variety of magnetic patterns (unpolarized,
ferromagnetic, and antiferromagnetic) accompanying the Mott and
superfluid phases.  Quantum Monte Carlo (QMC) calculations then are
used to explore the phase diagram more exactly.  In addition to
showing that many aspects of the interplay between superfluidity and
magnetism suggested by MFT persist, we also show that the order of the
chemical potential driven superfluid-Mott phase transition depends on
which Mott lobe is being considered, and even on whether commensurate
density is being approached from above or below.

High precision QMC work in two and three dimensions in the absence of
interconversion has previously demonstrated the existence of different
Mott and superfluid phases, distinguished by their patterns of charge
and spin order\cite{Capogrosso-Sansone2010}.  The possibility of
mixing `heavy' and `light' bosonic species (`mass imbalance')
introduced additional phenomena like ferromagnetic, phase separated
states, and `entropy squeezing', in which the heavy species is in a
Mott phase while the light species is superfluid and can act as a heat
reservoir to absorb entropy\cite{Hettiarachchilage2013}.  The effects
of interconversion on these phenomena is one of the topics of the
present work.

While we will explore here the phase diagrams for quite general values
of the kinetic and interaction energies, we note that the precise
quantitative form of the effective (pseudo) spin interaction potential
for ultracold bosonic and fermionic atoms can be computed using the
``degenerate internal state approximation''\cite{Santamore2011}.  At a
basic conceptual level, the coupling is similar to a true spin
interaction in which the magnetic field produced by one spin couples
to the second spin, but there are important differences.  One of these
is that, because the hyperfine states are not ``real spin'', they are
not generators of rotations, and hence there is no reason to expect an
isotropic (``Heisenberg-like'') form $J\,\hat\bfS_1 \cdot \hat\bfS_2$.  Instead,
the energy can be Ising or XY in character, and indeed the precise
form depends on the scattering lengths of binary atom-atom collisions
in the presence of an external field.

\section{The spin-$1/2$ model with near-neighbor spin interactions}

Here we are interested in the spin-$1/2$ Bose-Hubbard Hamiltonian\cite{Krutitsky:2004,Krutitsky:2005} 
with the spin interactions extended to near-neighbor sites

\begin{align}
  \hH = &-t\sum_{\ob{\bfi\bfj}\sigma} 
         \left( \hb_{\sigma\bfi}^\dagger \hb_{\sigma\bfj}^{\phantom{\dagger}} +
                \hb_{\sigma\bfj}^\dagger \hb_{\sigma\bfi}^{\phantom{\dagger}} \right)
       -\mu \sum_{\bfi,\sigma} \hn_{\sigma\bfi} + \frac{U_2}{4}\sum_\bfi \hat n^2_\bfi \nonumber\\
      &+ \frac{U_0}{2}\sum_{\bfi} \hn_\bfi(\hn_\bfi-1)  
       + U_2 \sum_\bfi\,\left( \hat S^x_\bfi \hat S^x_\bfi - \hat S^y_\bfi \hat S^y_\bfi - \hat S^z_\bfi \hat S^z_\bfi
                        \right)\nonumber\\
      &+ V   \sum_{\ob{\bfi\bfj}} 
             \left( \hat S^x_\bfi \hat S^x_\bfj - \hat S^y_\bfi \hat S^y_\bfj - \hat S^z_\bfi \hat S^z_\bfj
                        \right).
  \label{eq:BHspinHalfSU2}
\end{align}
In the above equation,   
$\hb_{\sigma\bfi}^\dagger$ ($\hb_{\sigma\bfi}^{\phantom{\dagger}}$) creates
(annihilates) a pseudo-spin $\sigma = \upa, \dna$ boson on site $\bfi$
of an $L\times L$ square lattice under periodic boundary conditions.
$\hn_\bfi=\hn_{\upa\bfi}+\hn_{\dna\bfi}$, and 
$\hat S^\alpha_\bfi$ ($\alpha=x,y,z$) is the spin operator defined as 
\be
  \hat S^\alpha_\bfi = \frac{1}{2}\sum_{\sigma\sigma'}\,\hat b^\dagger_{\sigma\bfi}\,\vec{\sigma}_{\sigma\sigma'}^\alpha\,
                                                        \hat b_{\sigma\bfi}^{\phantom{\dagger}},
  \label{eq:spinops}
\ee
where $\vec{\sigma}_{\sigma\sigma'}^\alpha$ are the Pauli matrices.
The parameters $t$ and $\mu$ correspond to the near-neighbor (NN)
hopping amplitude and chemical potential respectively. We use $t=1$ as
the unit of energy.  
$U_0$ is the contact interaction, while
$U_2$ and $V$ are on-site and NN spin-spin interactions respectively. 

Using the representation Eq.~(\ref{eq:spinops}) and taking the NN 
spin-spin interaction to be along the $z$-axis only\footnote{This Ising form corresponds to
a positive value of the inter-species scattering
length\cite{Timmermans1998}
}, 
we arrive at the following model Hamiltonian
\begin{align}
  \hH_z = &-t\sum_{\ob{\bfi\bfj}\sigma} 
         \left( \hb_{\sigma\bfi}^\dagger \hb_{\sigma\bfj}^{\phantom{\dagger}} +
                \hb_{\sigma\bfj}^\dagger \hb_{\sigma\bfi}^{\phantom{\dagger}} \right)
       -\mu \sum_{\bfi,\sigma} \hn_{\sigma\bfi} \nonumber\\
      &+ \frac{U_0}{2}\sum_{\bfi,\sigma} \hn_{\sigma\bfi}(\hn_{\sigma\bfi}-1)  
       + (U_0 + U_2)\sum_\bfi\,\hn_{\upa\bfi} \hn_{\dna\bfi}  \nonumber\\
      &+ \frac{U_2}{2}\sum_\bfi\left( \hb_{\upa\bfi}^\dagger \hb_{\upa\bfi}^\dagger 
                                      \hb_{\dna\bfi}^{\phantom{\dagger}} \hb_{\dna\bfi}^{\phantom{\dagger}} +
                                      \hb_{\dna\bfi}^\dagger \hb_{\dna\bfi}^\dagger 
                                      \hb_{\upa\bfi}^{\phantom{\dagger}} \hb_{\upa\bfi}^{\phantom{\dagger}} 
                              \right) \nonumber\\
      & -V\sum_{\ob{\bfi\bfj}} \hS^z_\bfi \hS^z_\bfj.
  \label{eq:BHspinHalf}
\end{align}
It can be seen that the onsite spin coupling $U_2$ has two
roles. First, it shifts the strength of the contact interaction
between opposite spins $n_{\upa\bfi} n_{\dna\bfi}$. Second, $U_2$ is
also the matrix element of the conversion process which turns two
identical bosons into the opposite spin species when they meet at the
same site.

Eq.~(\ref{eq:BHspinHalf}) is the Hamiltonian that will be studied in
this work. It will be solved using MFT and exact stochastic Green function
(SGF) quantum Monte Carlo technique.
While the SGF QMC method can treat the contact spin interactions {\it or}
the NN spin-spin couplings separately, a sign problem 
would arise if both 
terms were retained due to the presence of interconversion matrix elements in both.
 For this technical reason, we drop the conversion matrix elements of $V$ terms 
 in Eq.~(\ref{eq:BHspinHalfSU2}) 
 and study Eq.~(\ref{eq:BHspinHalf}).
The retention of the $z$-axis term gives important insights 
into the effects of the NN spin-spin interactions, in the same spirit that the 
$t$-$J_z$ Hamiltonian provides initial clues into the more general rotationally 
invariant $t$-$J$ model.

We focus on the case where $U_2>0$ and $V<0$, i.e., antiferromagnetic NN spin couplings.  
In general, the value of $U_0$, $U_2$ and $V$ will depend on details of the 
system (for example, scattering length between the atoms, polarization of the 
laser waves forming the optical lattice, detuning from the internal atomic 
transition etc.\cite{Krutitsky:2004}). Here we adopt the parameter regime
studied in Ref.~\onlinecite{Krutitsky:2004} where,
based on known values of $^{87}$Rb and $^{23}$Na scattering lengths
and on laser wavelengths corresponding to the $D_1$ resonance,
$U_2$ is typically an
order of magnitude or more smaller than $U_0$. In the current work, we take
the value $U_2/U_0 = 0.1$. 
For the NN coupling $V$, because interaction strength typically decreases with
distance, we assume the value $|V|/U_0 < 0.1$.

When $V=0$ in Eq.~(\ref{eq:BHspinHalf}), the system was
studied extensively by MFT\cite{Krutitsky:2004,Krutitsky:2005} and QMC
methods in one and two dimensions\cite{Larent:2010,Larent:2011} . In
2D and $U_2 >0$\footnote{From now on when $U_2$ is referenced, it is
meant to be the $U_2$ in front of the contact interaction term in
Eq.~(\ref{eq:BHspinHalf})}, the ground state of the Hamiltonian
features three phases: a ferromagnetic superfluid (FMSF), an
unpolarized Mott insulator (MI) at even commensurate densities, and a
ferromagnetic Mott phase at odd commensurate fillings.  For
negative $U_2$, it was found that the ground state never
polarizes\cite{Larent:2010,Larent:2011}.

\section{Mean field theory}

\subsection{Decoupling Mean Field theory}

The mean-field scheme employed in the present work is developed in
Ref.~\onlinecite{Sheshadri:1993,Oosten:2001}.  The method is based on
rewriting the Hamiltonian as a sum over local terms that can be solved
exactly for a fixed number of bosons.  To incorporate the hopping terms,
one introduces uniform SF order parameters
$\ob{\hb_{\sigma\bfi}^\dagger} = \ob{\hb_{\sigma\bfi}^{\phantom{\dagger}}} = \psi_\sigma$.
Since we are interested in equilibrium states, the order parameters
$\psi_\sigma$ can be chosen to be real. Using this ansatz, the kinetic
energy terms, which are non-diagonal in boson creation and destruction
operators, are decoupled as
\begin{align}
  \hb_{\sigma\bfi}^\dagger \hb_{\sigma\bfj}^{\phantom{\dagger}} 
   = & \,(\hb_{\sigma\bfi}^\dagger - \ob{\hb_{\sigma\bfi}^\dagger})
       (\hb_{\sigma\bfj}^{\phantom{\dagger}} - \ob{\hb_{\sigma\bfj}^{\phantom{\dagger}}}) \nonumber\\
     & + \ob{\hb_{\sigma\bfi}^\dagger} \hb_{\sigma\bfj}^{\phantom{\dagger}} + \hb_{\sigma\bfi}^\dagger \ob{\hb_{\sigma\bfj}^{\phantom{\dagger}}}
       - \ob{\hb_{\sigma\bfi}^\dagger} \ob{\hb_{\sigma\bfj}^{\phantom{\dagger}}} \nonumber\\     
    \approx & \, (\hb_{\sigma\bfi}^\dagger + \hb_{\sigma\bfj}^{\phantom{\dagger}})\psi_\sigma + \psi_\sigma^2,
\end{align}
where in the last line we have dropped the terms that
have products of fluctuations in bosonic operators on different sites.

To treat the NN spin interactions in the same decoupling scheme, we
decompose the square lattice into two disjoint sublattices $A$ and $B$
and introduce real magnetic order parameters $\ob{\hS_A^z}$,
$\ob{\hS_B^z}$ on sublattice $A$ and $B$ respectively.  Under the MF
approximation, the spin-spin interaction term now becomes
\begin{align}
  |V|\sum_{\ob{\bfi\bfj}} \hS_\bfi^z \hS_\bfj^z
     \approx   &\,\, z_c |V|\sum_{\bfi\in A} 
               \left(\, \hS_\bfi^z\ob{\hS_B^z} - \frac 1 2 \ob{\hS_A^z}\ob{\hS_B^z} \,\right) \nonumber\\
               & + z_c |V| \sum_{\bfj\in B} 
               \left(\,\hS_\bfj^z\ob{\hS_A^z} - \frac 1 2 \ob{\hS_A^z}\ob{\hS_B^z} \,\right).
\end{align}
As before, in this form, terms that describe products of fluctuations in
$\hS_A^z$ and $\hS_B^z$ are ignored.  $z_c=4$ is the coordination number
of the square lattice. We have also assumed that the magnetic order
parameter on each sublattice is uniform.  With these approximations, the
final MF Hamiltonian becomes two coupled local ones for sublattices $A$
and $B$:
\begin{align}
 \hH_\ell = & - z_c t \sum_\sigma (\hb_{\sigma\ell}^\dagger + \hb_{\sigma\ell})\psi_{\bar\sigma\ell}
         + z_c t \sum_\sigma \psi_{\sigma\ell}\psi_{\bar\sigma\ell} \nonumber\\
       & + \frac{U_0}{2} \hn_{\upa \ell} (\hn_{\upa \ell}-1) + \frac{U_0}{2} \hn_{\dna\ell} (\hn_{\dna\ell}-1) \nonumber\\
       & + \frac{U_2}{2} \left( \hb_{\upa \ell}^\dagger \hb_{\upa \ell}^\dagger 
                                \hb_{\ell \dna}^{\phantom{\dagger}} \hb_{\dna\ell}^{\phantom{\dagger}} + h.c. \right) 
          + (U_0 + U_2) \hn_{\upa \ell} \hn_{\dna\ell} \nonumber\\
       &  - \mu (\hn_{\upa \ell} + \hn_{\dna\ell}) 
         + z_c |V| \hS_\ell^z\ob{\hS_{\bar\ell}^z} 
         - \frac{z_c |V|}{2} \ob{\hS_\ell^z}\ob{\hS_{\bar\ell}^z}, \label{eq:MFspinhalf}
\end{align}
where $\ell=A$, $B$ ($\bar\ell=B$, $A$), and $\sigma=\,\upa,\dna$
($\bar\sigma=\,\dna,\upa$).  The coupled Hamiltonians are solved at
zero temperature by iteration. Starting with an initial guess of order
parameters $\psi_{\sigma\ell}$ and $\ob{\hS^z_\ell}$, $\hH_A$ and
$\hH_B$ can be diagonalized numerically within the bosonic occupation
number basis $\{\ket{n_{\upa \ell}, n_{\dna\ell}}\}$ truncated at
$n_{\sigma\ell} \leq N_{\max}$. Order parameters are then updated with
respect to the new MF ground state.  This procedure is repeated until
$\psi_{\sigma\ell}$, $\ob{\hS^z_\ell}$ and the ground state energy
$E_g$ are converged. We typically choose $N_{\max}=14$ to ensure that
convergence is independent of $N_{\max}$.  Multiple initial
configurations are also used to verify that the converged MF solution
do not depend on initial conditions.  We benchmark our MF program by
computing the phase diagram of Eq.~(\ref{eq:MFspinhalf}) with $V=0$.
The results are in agreement with previously published
data\cite{Larent:2010}.

\begin{figure}
\includegraphics[scale=0.37]{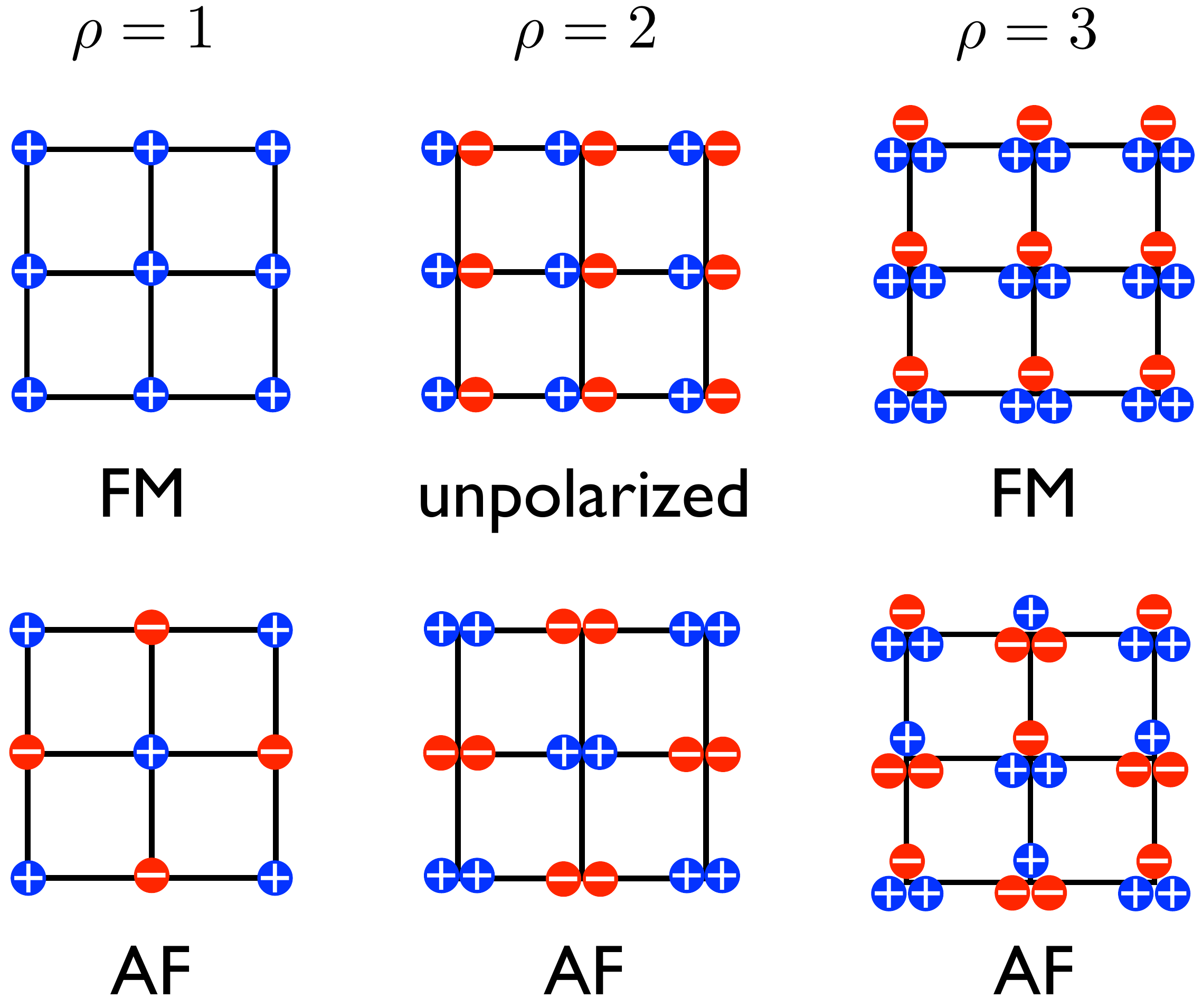}
\caption{(Color online)
Schematic diagram of ferromagnetic, antiferromagnetic, and unpolarized states
at different commensurate fillings. Blue ($+$) and red ($-$) circles represent 
spin-up and spin-down components respectively.
} 
\label{fig:Cartoon}
\end{figure}

Different MF phases are classified by the corresponding order
parameters.  For example, a superfluid is characterized by finite
total superfluid density \be \rho_{s,\ell} = \rho_{s,\upa \ell} +
\rho_{s,\dna\ell} = \psi_{\upa \ell}^2 + \psi_{\dna\ell}^2.  \ee The
Mott insulator, on the other hand, is defined by zero superfluid
density $\rho_{s,\ell}=0$ and zero compressibility
$\partial\rho_\ell/\partial\mu=0$, where \be \rho_\ell = \rho_{\upa
  \ell} + \rho_{\dna \ell}.  \ee To examine magnetic order, we compute
the expectation value of $\hS^z$ with respect to the converged MF
solution. In a Mott phase, this is
\begin{align}
  S^z_\ell &= \frac 1 2 \ob{\hn_{\upa \ell} - \hn_{\dna\ell}},
  \label{eq:SzMI}
\end{align}
where $\hn_{\sigma\ell}$ is the density operator.
In principle one can use Eq.~(\ref{eq:SzMI}) in the SF phase, 
and the conclusion should remain the same.
Here we follow the convention in Ref.~\onlinecite{Ho:1998,Krutitsky:2004} 
and compute the magnetization in the SF phase defined as
\be
  S^z_{s,\ell} = \frac 1 2 \frac{\psi_{\upa \ell}^2 -
    \psi_{\dna\ell}^2}{\rho_{s,\ell}}, 
\ee
which merely measures the SF population difference between the two spin
components.

Figure~\ref{fig:Cartoon} summarizes schematically the possible
magnetic structures at three commensurate fillings. For example, the
state (SF or MI) is ferromagnetic (FM) if one of the spin components
dominates the population throughout the lattice. An unpolarized state
has both spin components equally occupied on every lattice site. An
antiferromagnetic (AF) state is realized when sublattices $A$ and $B$
are dominated by different spin species.

\subsection{Mean Field Results}

\begin{figure}
\includegraphics[scale=0.425]{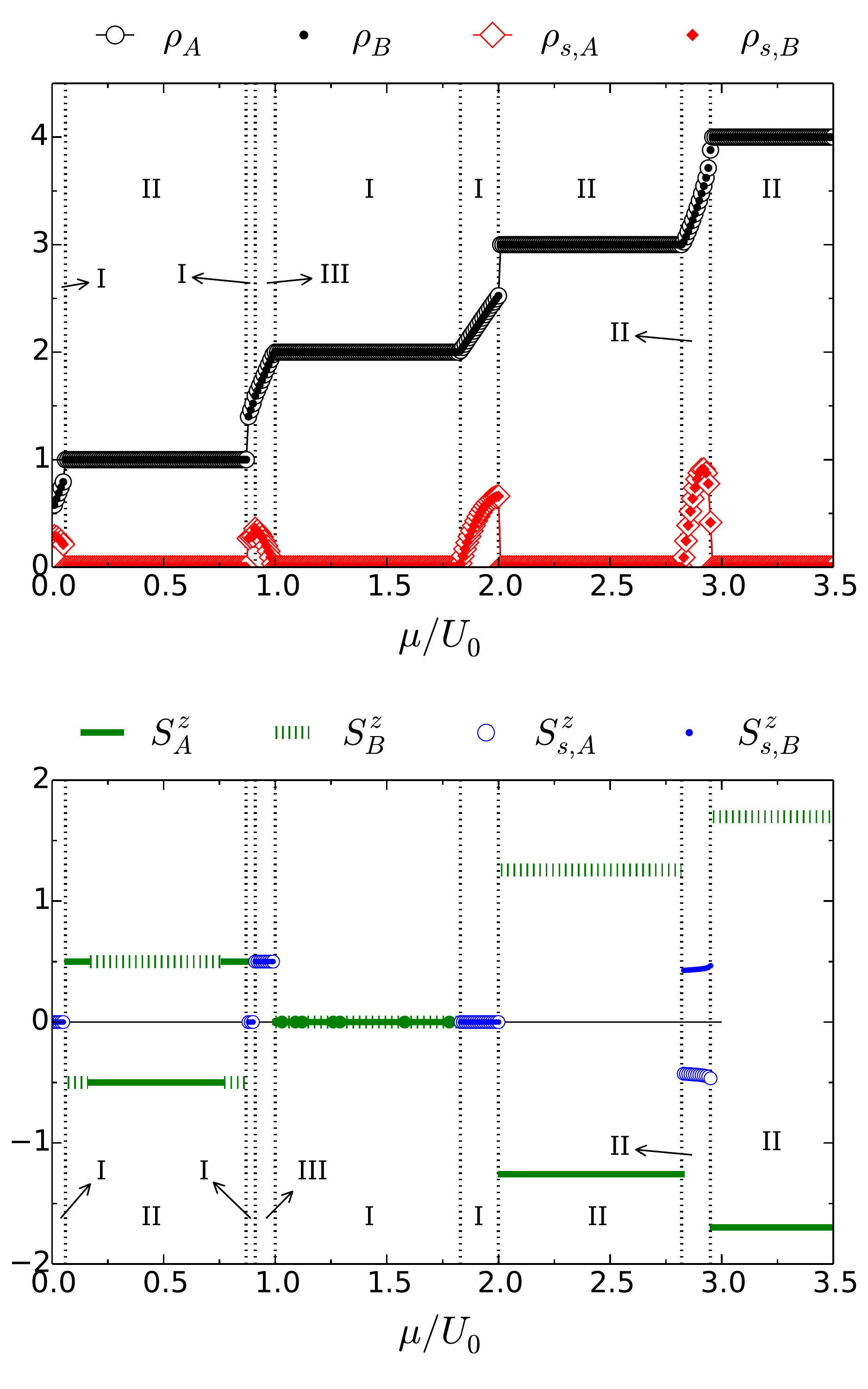}
\caption{(Color online)
Features of the MF ground state as a function of chemical potential
$\mu/U_0$ at $t/U_0=0.02$, $U_2/U_0=0.1$ and $|V|/U_0=0.02$.  Total
particle and total superfluid densities on $\ell=A$ and $B$ sublattices
are plotted in the top panel. Magnetic order parameters $S^z_\ell$ and
$S^z_{s,\ell}$ are shown in the bottom panel. In both figures, the
vertical dashed lines divide the phases into various zones based on
their magnetic structures labeled by the roman numerals: I --
unpolarized, II -- AF, and III -- FM.  Note that 
the changes in sign of $S^z_A$ ($S^z_B$) 
in the first Mott plateau are a trivial reflection of 
the degeneracy of the two MF solutions.
}
\label{fig:MFcut}
\end{figure}

Properties of the MF ground state are shown in Fig.~\ref{fig:MFcut}
for $t/U_0=0.02$, $U_2/U_0=0.1$, and $|V|/U_0=0.02$. Total particle and
SF densities are plotted in the upper figure as functions of
$\mu/U_0$.  The density develops three well defined plateaux at
$\rho=1$, 2, and 3.  These plateaux correspond to MIs because the
compressibility $\partial\rho/\partial\mu=0$ and the SF density also
vanishes. The SF phase resides in between the Mott insulators.  It can
be seen that $\rho_\ell$ and $\rho_{s,\ell}$ change
discontinuously when one enters and leaves the first Mott
plateau. This is the signature of a first order phase 
transition. Likewise, the transition is also first order as one enters
the $\rho=3$ MI from below. On the other hand, $\rho_\ell$ and
$\rho_{s,\ell}$ change continuously on both sides of the second Mott
plateau, indicating that the MI-SF transition is second order.

The lower panel of Fig.~\ref{fig:MFcut} summarizes MF magnetic
structures for $t/U_0=0.02$, $U_2/U_0=0.1$, and $|V|/U_0=0.02$. Within
the $\rho=1$, 3, and 4 plateaux, the magnetic order parameter on
sublattices $A$ and $B$ are equal but have opposite signs $S^z_A = -
S^z_B$.  This shows that these MIs are antiferromagnetic. In contrast,
the second Mott insulating region is non-magnetic.  In the SF region,
magnetic properties are plotted by blue symbols (dots and empty
circles).  The SF between the $\rho=1$ and $\rho=2$ Mott regions has
two different magnetic natures: unpolarized and fully polarized. The
transition between them is first order. This is also indicated in the
upper panel by a discontinuity in $\rho_{s,\ell}$.  Most
interestingly, the SF above the $\rho=3$ Mott region shows
antiferromagnetic structure.

\begin{figure}
\includegraphics[scale=0.43]{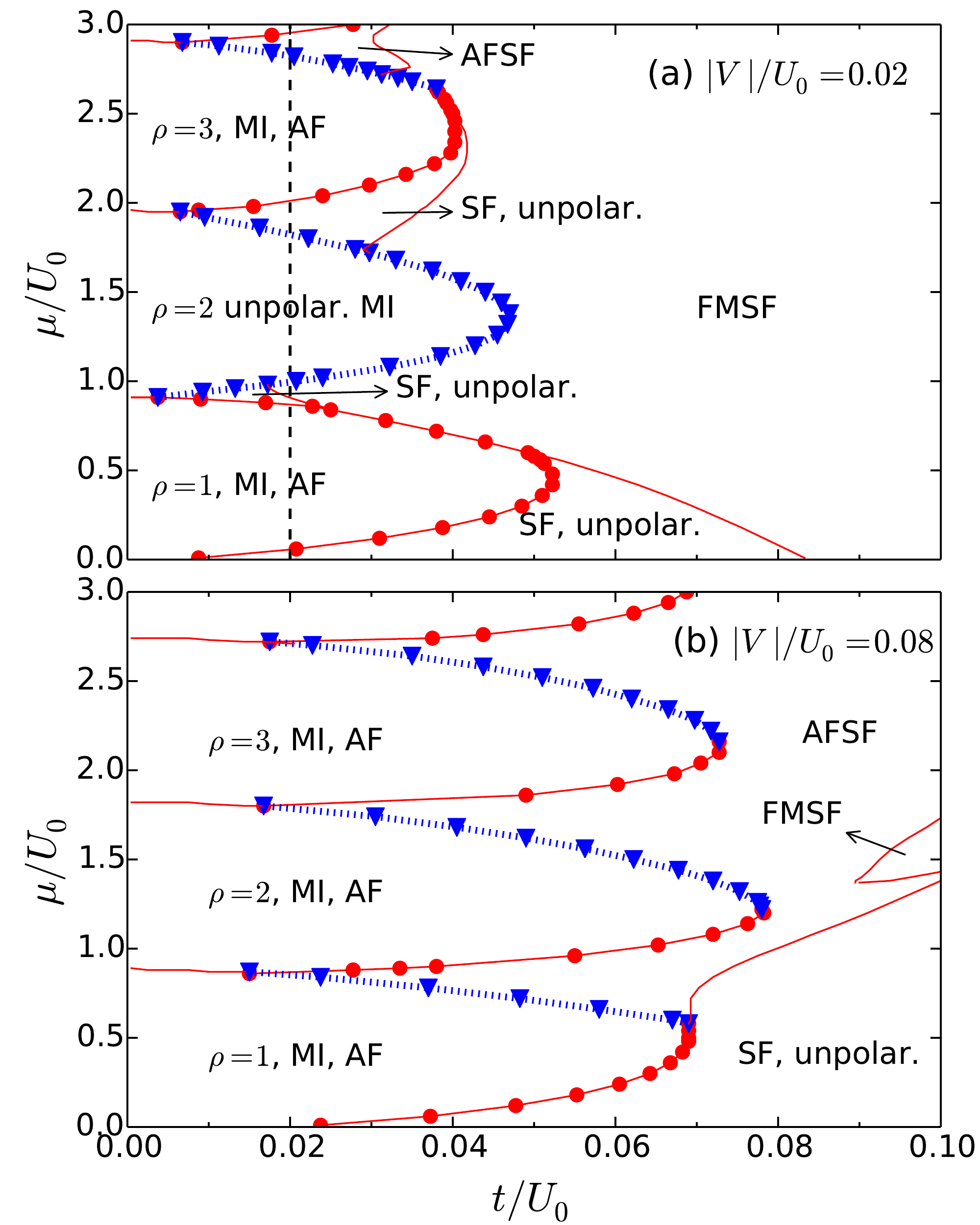}
\caption{(Color online) MF phase diagrams obtained by solving the
  coupled Hamiltonian Eq.~(\ref{eq:MFspinhalf}) for $U_2/U_0=0.1$. The
  NN spin interaction strength $|V|/U_0$ is (a) 0.02 and (b) 0.08.
  First order and continuous phase transitions are represented by red
  (solid) and blue (dashed) curves respectively. ``AF'', ``FM'', and
  ``unpolar'' indicate magnetic structures.  At $|V|/U_0=0.02$, an AFSF
  phase emerges between the $\rho=3$ and $4$ Mott lobes. With
  increasing $|V|/U_0$, the AFSF region expands significantly.
  Moreover, the $\rho=2$ lobe changes its nature from being
  non-magnetic at $|V|/U_0=0.02$ to antiferromagnetic at $|V|/U_0=0.08$.
  In panel (a), the vertical dashed line indicates the location of
  $t/U_0$ where Fig.~\ref{fig:MFcut} is plotted.  }
\label{fig:MFPD}
\end{figure}

By carrying out the self-consistent MF calculation at different
$t/U_0$ (or $\mu/U_0$) values, the $\mu$-$t$ phase diagram can be
constructed.  Results for $|V|/U_0=0.02$ and 0.08 are plotted in
Fig.~\ref{fig:MFPD}.  Here red (solid) and blue (dashed) curves
represent first and second order phase transitions respectively.
Comparing with the $V=0$ phase diagram\cite{Larent:2010}, there are
several notable changes due to the presence of NN spin-spin couplings.

The magnetic structure of the first and third Mott lobes changes from
being ferromagnetic to antiferromagnetic. At $\rho=1$ or 3, one of the
spin components dominates the population. As a result, the MF ground
state energy can be lowered by forming an AF pattern. At $\rho=2$, on
the other hand, the onsite coupling term can be effectively avoided by
equally populating both spin species on every site if $|V|/U_0=0.02$ is
small. By raising $|V|/U_0$ to 0.08, the second lobe also becomes
antiferromagnetic.  This is because the energy gained by forming an AF
state compensates the energy cost of onsite coupling terms at large
$|V|/U_0$ values.

When $V=0$, the MI-SF phase transition is continuous except for the
tip of the second Mott lobe. The transition is known to be first order
for $0 < U_2/U_0 \lesssim 0.25$.\cite{Larent:2010} Here we find that
the transition becomes first order due to the change of magnetic
property in the $\rho=1$ and 3 Mott lobes. Note that above the third
lobe, the antiferromagnetic Mott insulator to antiferromagnetic
superfluid (AFSF) transition remains continuous. At $|V|/U_0=0.08$, the
bottom half of the phase boundary enclosing the Mott insulators is
first order; while the upper half becomes continuous.

Regarding the magnetic structure of the SF phase, it was found that
the SF is always polarized if $V=0$\cite{Larent:2010}. With the
presence of NN spin-spin couplings, an unpolarized SF emerges near the
Mott lobes, particularly at small $t/U_0$ values. An exception to this
observation is found at $\rho > 3$ where an AFSF phase occupies the
region between the third and fourth MIs. At $|V|/U_0=0.08$, the AFSF
region expands dramatically to large hopping regions, and to chemical
potential values as low as $\mu/U_0 \sim 0.6$. This AFSF is a
supersolid phase since it exhibits simultaneous diagonal and
off-diagonal long range order.

Recall that in the original Bose-Hubbard model\cite{Fisher:1989} or in 
the case $V=0$ in Eq.~(\ref{eq:MFspinhalf})\cite{Larent:2010,Larent:2011},
the SF phase extends all the way to $t/U_0=0$. 
Fig.~\ref{fig:MFPD} shows that this is no longer the case when $V$ is
turned on. The system undergoes a series of first-order transition
between MIs at small $t/U_0$.

\section{Exact Quantum Monte Carlo Study}

In this section, we solve the model Eq.~(\ref{eq:BHspinHalf}) exactly
on finite lattices by using Stochastic Green Function (SGF)
QMC\cite{DirectedSGF}.  The SGF method is a finite-temperature
continuous time QMC technique that can be formulated in either the
canonical or grand canonical ensembles. The SGF algorithm can solve a
large class of lattice Hamiltonians that can be written as $\hH=\hV -
\hT$, where $\hV$ is diagonal in the Fock basis (subject to the model
type) and $\hT$ has only positive elements\cite{DirectedSGF}. The
technique has also been applied to the $V=0$ case of
Eq.~(\ref{eq:BHspinHalf}) in one and two
dimensions\cite{Larent:2010,Larent:2011}.

In our simulations, the temperature is set at $\beta t = 2L$ for a
lattice with linear dimension $L$. The chosen temperature is typically
low enough to ensure that the results are converged to the ground
state limit. In some cases, we select $\beta t=4L$ to reach
convergence.  We benchmarked the SGF algorithm by comparing with exact
diagonalization data for a small cluster. The SGF and exact results
are in agreement within statistical errors.

\subsection{Phase diagram}

\begin{figure}
\includegraphics[scale=0.39]{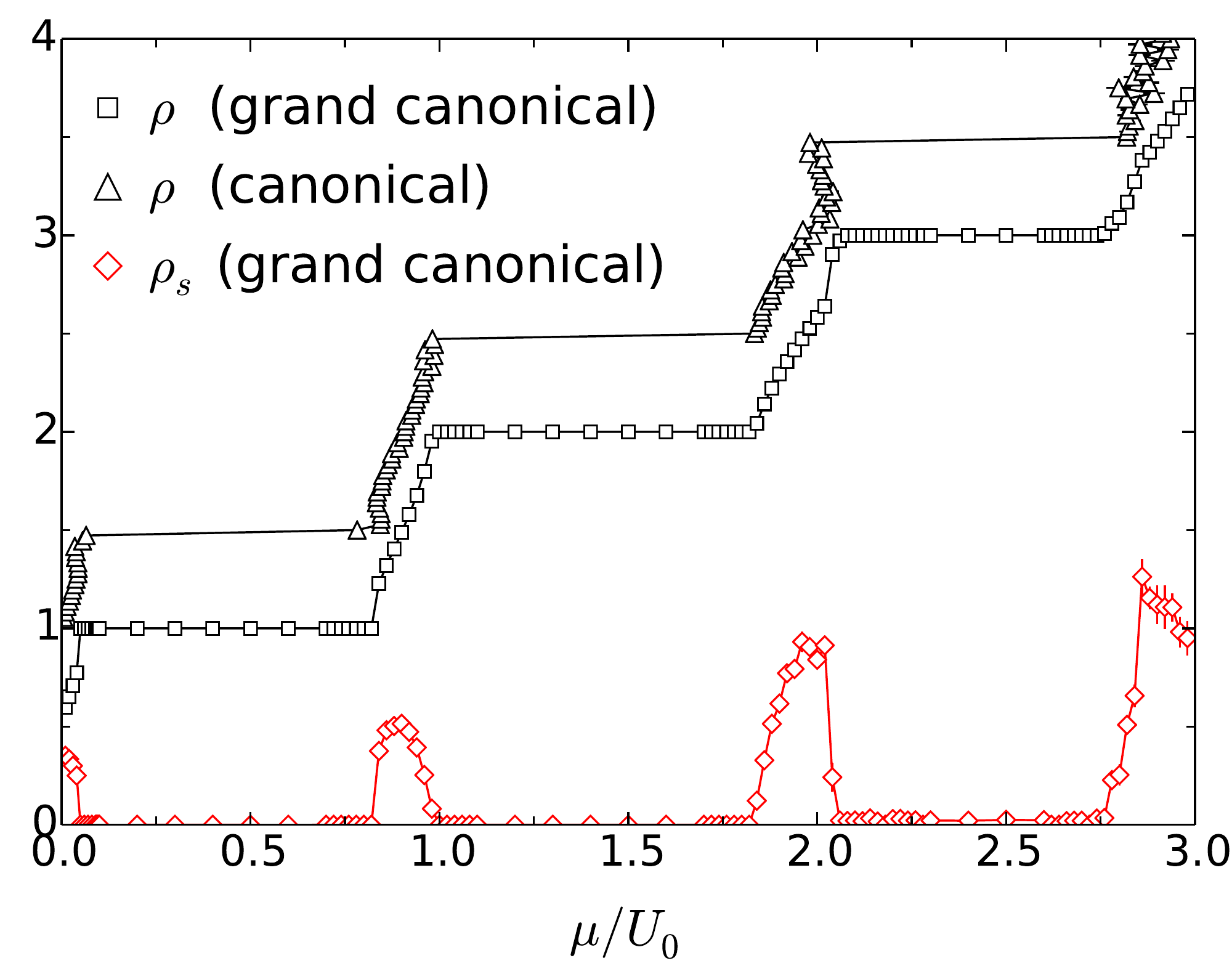}
\caption{(Color online) Particle density $\rho$ (square and triangle)
  and SF density $\rho_s$ (diamond) versus chemical potential computed
  using the SGF QMC method for $t/U_0=0.02$.  For comparison, we
  have implemented canonical and grand canonical ensembles in the
  simulations.  Both data sets are acquired on a $L=6$ lattice at
  temperature $\beta=12$. Conversion term and NN spin interactions are
  $U_2/U_0=0.1$ and $|V|/U_0=0.02$ respectively.  The canonical ensemble
  results are shifted upward by 0.5 for clarity.  The signature of
  first order transition (discontinuous jump in $\rho$ and $\rho_s$ in
  the grand canonical ensemble curve, and negative compressibility
  $\partial\rho/\partial\mu < 0$ in the canonical ensemble data) can
  be seen at both ends of the first Mott lobe and the bottom
  boundary of the third lobe.  }
\label{fig:QMCcut}
\end{figure}

\begin{figure}
\includegraphics[scale=0.42]{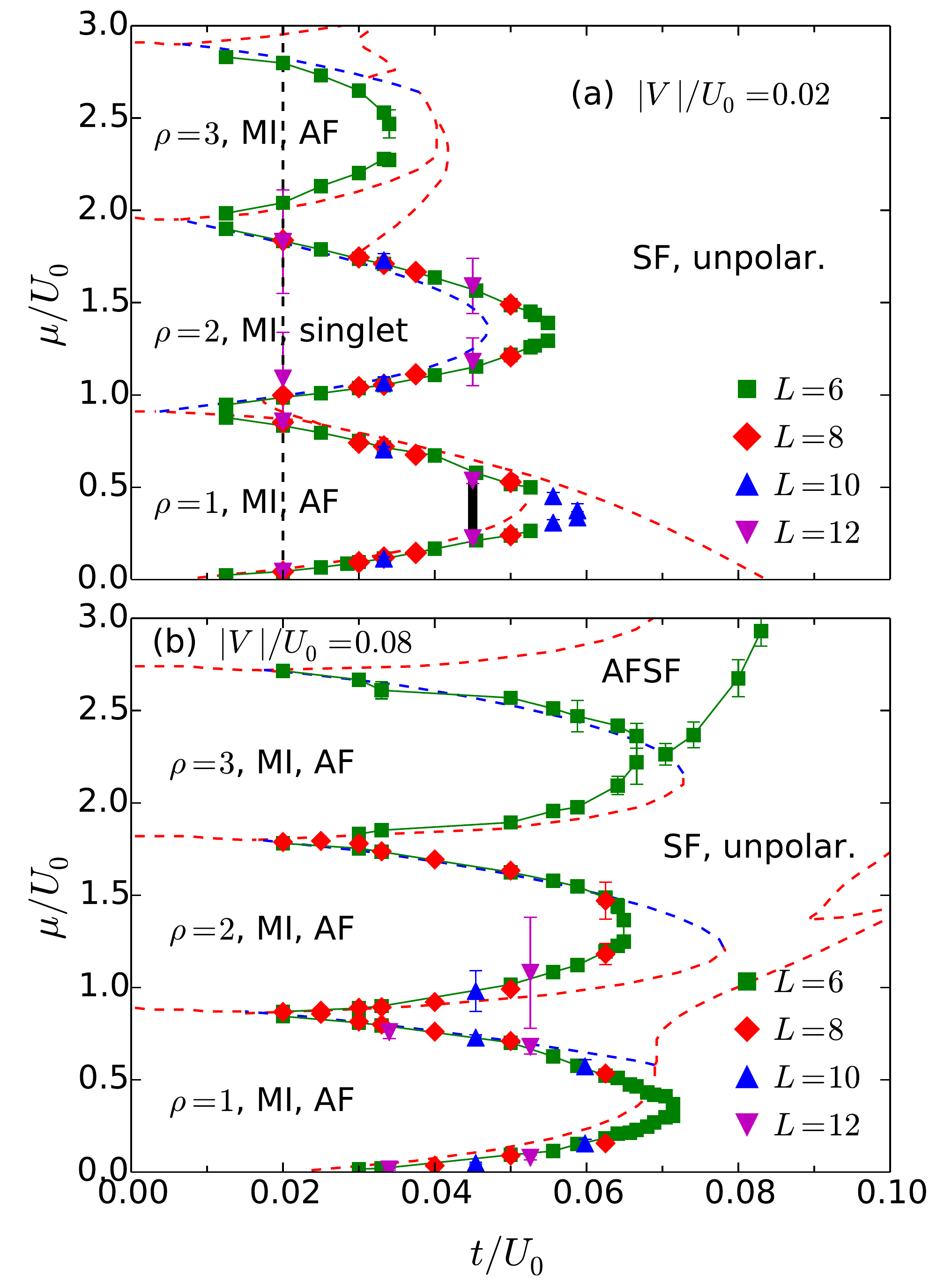}
\caption{(Color online) Exact phase diagram of the Hamiltonian
  Eq.~(\ref{eq:BHspinHalf}) obtained using the SGF QMC technique on
  $L\times L $ lattices at temperature $\beta=2L$.  The onsite spin
  coupling is fixed at $U_2/U_0=0.1$. The NN spin-spin interaction
  $|V|/U_0$ is set at (a) 0.02 and (b) 0.08.  Statistical uncertainties
  are typically smaller than the symbol size.  The vertical dashed
  line is the location where Fig.~\ref{fig:QMCcut} is plotted.
  Mean-field phase boundaries are also presented (dashed curves) for
  comparison.  In (a), the vertical heavy line in the $\rho=1$
  Mott lobe near the tip is the location of magnetic transition 
  (c.f. Fig.~\ref{fig:SafSF.vs.t_rho1}).  In (b), 
  AFSF is predicted to exist in the region above the $\rho=3$ MI.
}
\label{fig:QMCPD}
\end{figure}

To construct the exact phase diagram, we compute the total particle
density $\rho$ and SF density $\rho_s$ as functions of chemical
potential. In canonical ensemble SGF simulations, the total particle
number $N$ is fixed, and we derive the chemical potential via \be
\mu(N) = E(N+1) - E(N), \ee where $E(N)$ is the total energy of $N$
bosons on an $L\times L$ lattice.  To access the SF density, we use
the formula proposed by Pollock and Ceperley\cite{Pollock1987}, which
relates $\rho_s$ to the winding number $W$. However, due to the
conversion term in the Hamiltonian Eq.~(\ref{eq:BHspinHalf}), the
numbers of spin $\upa$ and $\dna$ bosons are not conserved
individually. As a consequence, the relevant winding number should
take into account both spin components\cite{Eckholt2010} and $\rho_s$
is given by the following formula
\be \rho_s = \frac{\ob{(W_\upa+W_\dna)^2}}{2dt\beta
  L^{d-2}}, 
\ee 
where $d=2$ is the dimensionality, $t$ is the
hopping amplitude, and $W_\upa$ and $W_\dna$ are the winding numbers
of spin $\upa$ and $\dna$ bosons respectively.

Figure~\ref{fig:QMCcut} shows QMC results for $\rho$ and $\rho_s$
versus $\mu/U_0$ on the $L=6$ lattice with $t/U_0=0.02$,
$U_2/U_0=0.1$, and $|V|/U_0=0.02$.  We compare total densities
$\rho(\mu)$ measured using both grand canonical (square) and canonical
(triangle) ensembles.  Three plateaux can be observed at
commensurate fillings.  Since the compressibility
$\partial\rho/\partial\mu$ and superfluid density $\rho_s$ vanish in
the plateaux, these regions represent Mott insulators.  In between the
Mott insulators there is a SF with $\rho_s\neq 0$.  The agreement of
the data for different ensembles acts both as a check of our codes and
also as an assessment of finite size effects, since equivalence is
expected only for sufficiently large lattices.

In Fig.~\ref{fig:QMCcut}, the grand canonical ensemble particle
density has a discontinuous jump when one enters or leaves the first
Mott region.  These jumps show that the MI-SF transition is first
order. This is confirmed by the canonical ensemble data which clearly
indicates negative compressibility $\partial\rho/\partial\mu <0$ in
the same region. At the same time, the SF density shows a
discontinuous jump.  Likewise, the MI-SF transition near $\mu/U_0\sim
2.1$ is first order.  The transition at $\rho=2$ is second order as
both quantities $\rho$ and $\rho_s$ change continuously (within the
resolution of our $\mu/U_0$ grid) as a function of the chemical
potential. MF predictions at $t/U_0=0.02$ (cf. Fig.~\ref{fig:MFcut})
are consistent with these exact QMC results.

The QMC phase diagram is shown in Fig.~\ref{fig:QMCPD} for (a)
$|V|/U_0=0.02$ and (b) $|V|/U_0=0.08$. The onsite spin coupling strength
is $U_2/U_0=0.1$ in both cases. The corresponding MF phase boundaries
(dashed curves) are also plotted for comparison. The QMC data are
shown for $\mu/U_0 \lesssim 2.75$ as it becomes increasingly difficult
to reduce statistical errors for simulation at large chemical
potential values. System sizes $L=6$, 8, 10, and 12 are used, with
little variation evident on the QMC phase boundaries.  Overall, the
QMC and MF phase boundaries are in good agreement, especially at small
$t/U_0$ where the MF assumption works well. The deviation between the
two approaches increases as one moves toward the tips of the Mott
lobes where quantum fluctuations are large.  
  Interestingly, at $/U_0=0.02$ and near $t/U_0\sim 0.045$, our QMC
  data reveal a magnetic phase transition inside the first MI.  The
  transition is indicated by a thick black line in
  Fig.~\ref{fig:QMCPD} and will be discussed in the next subsection.
  At $|V|/U_0=0.08$, the QMC Mott insulating regions expand, which is
consistent qualitatively with MF results.  At small $t/U_0$ values,
our QMC data also suggest the existence of a direct first-order MI-MI
transition at both $|V|/U_0=0.02$ and 0.08, confirming the MF
predictions.

\begin{figure}
\includegraphics[scale=0.36]{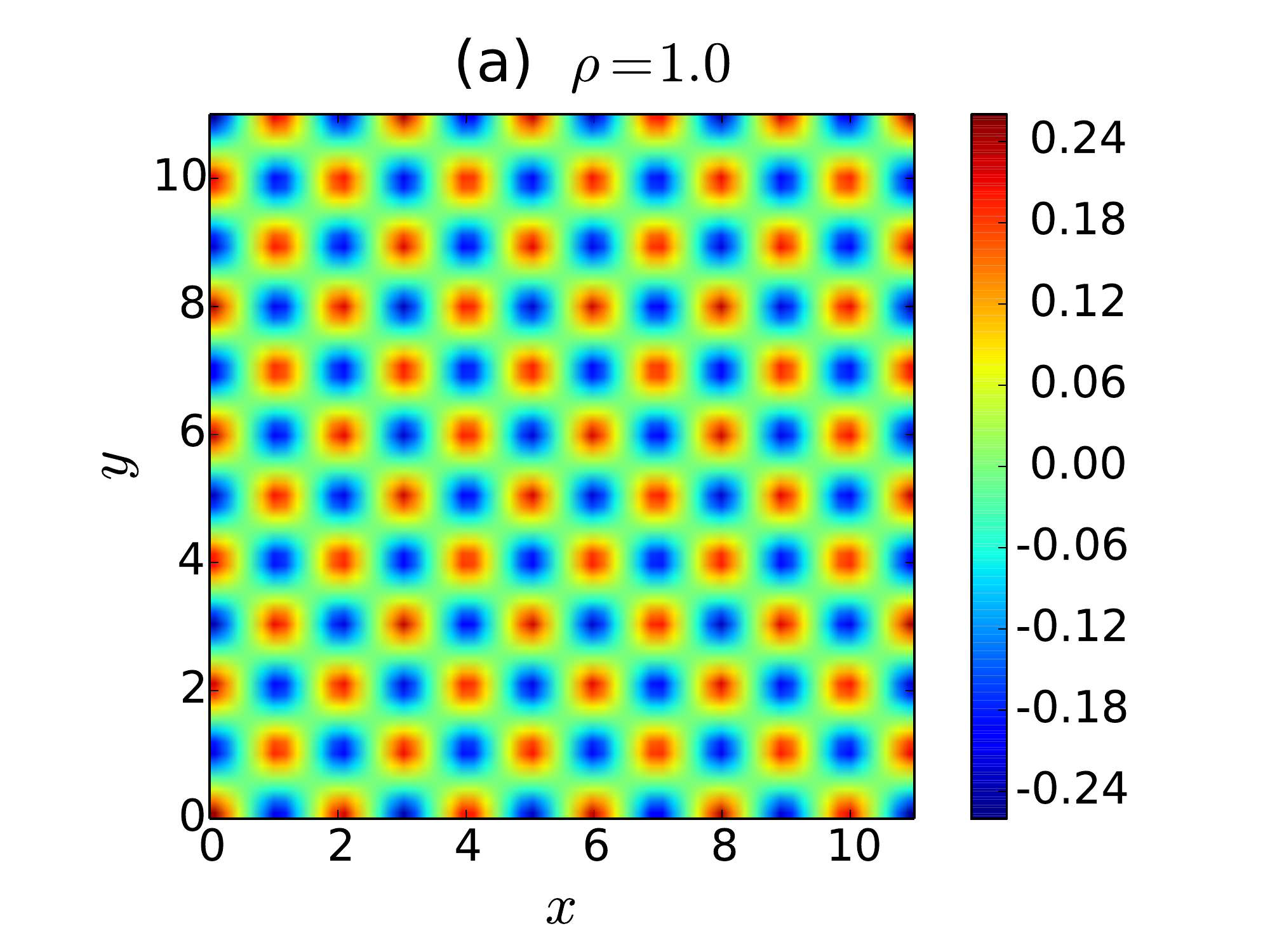}\\
\includegraphics[scale=0.36]{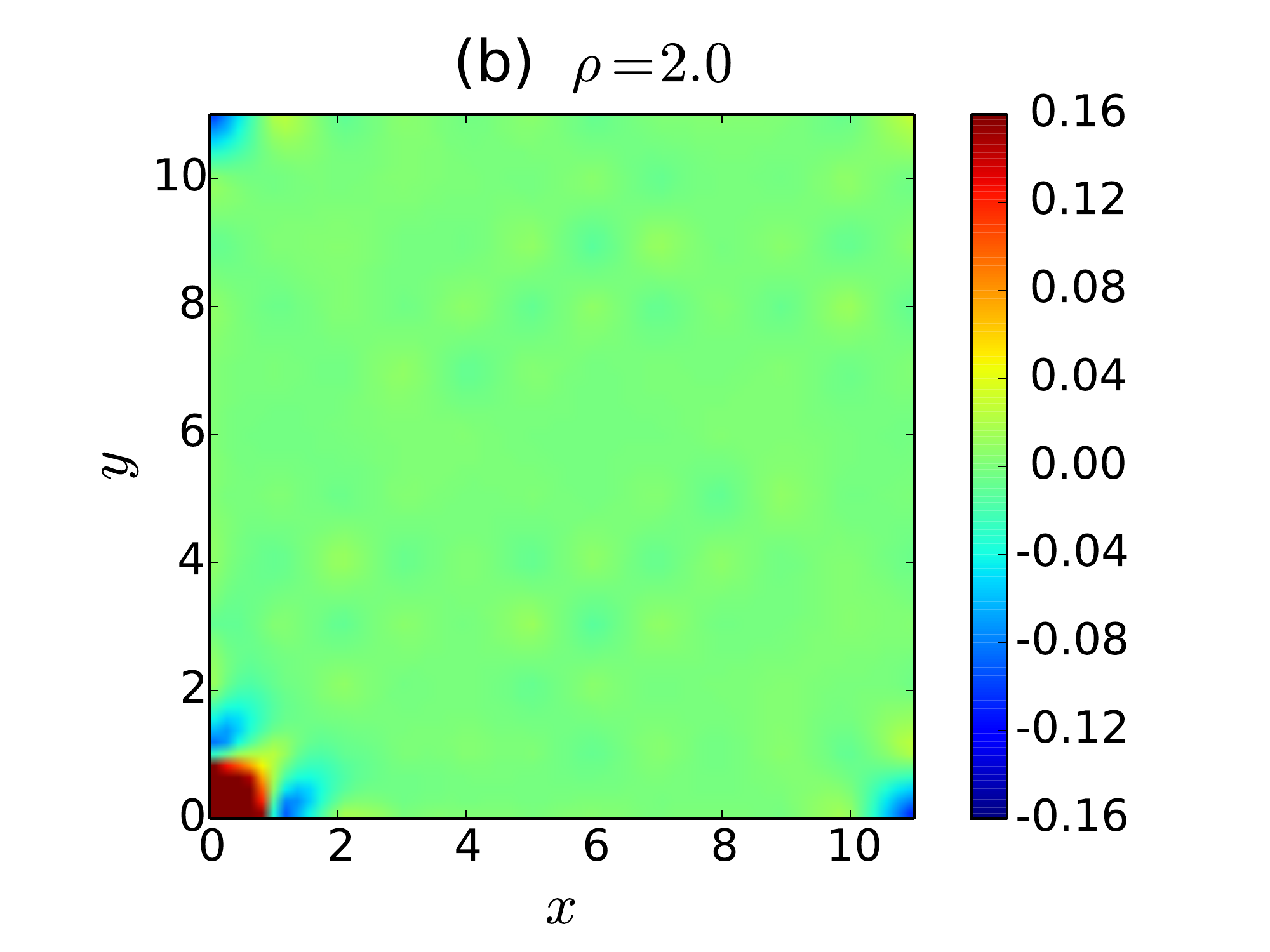}
\caption{(Color online) Spin-spin correlation function $C^{zz}(\bfr)$
  for the (a) $\rho=1$ and (b) $\rho=2$ Mott insulators. The data are
  measured on an $L=12$ lattice for $t/U_0=0.02$, $U_2/U_0=0.1$, and
  $|V|/U_0=0.02$. The data show that the first lobe has AF order while
  the second lobe is non-magnetic.  }
\label{fig:SpinCorr}
\end{figure}

\subsection{Magnetic properties of the Mott lobes}

\begin{figure}
\includegraphics[scale=0.44]{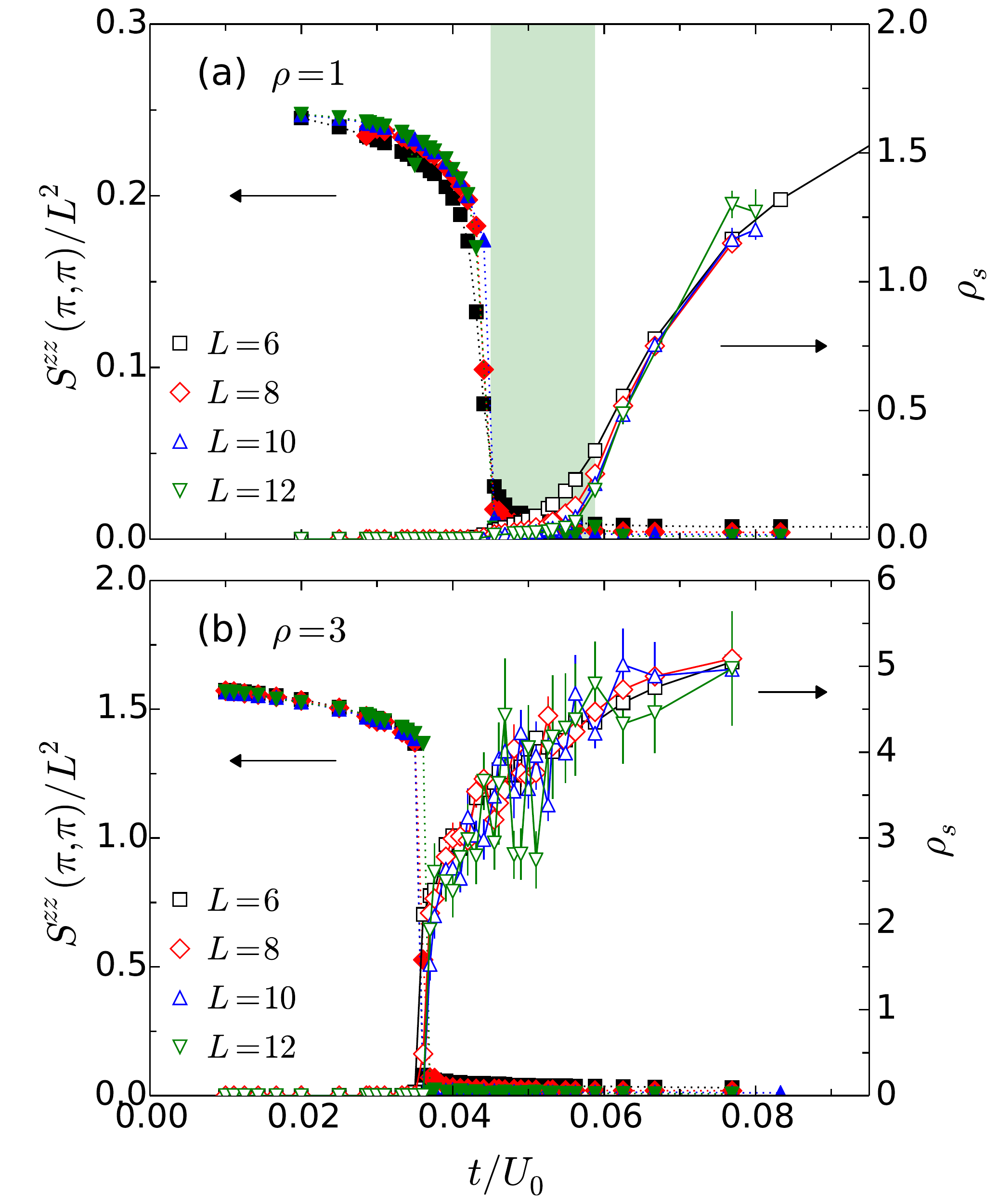}
\caption{(Color online) Normalized antiferromagnetic spin structure
  factor $S^{zz}(\pi,\pi)/L^2$ and SF density $\rho_s$ as functions of
  hopping amplitude $t/U_0$. Here $U_2/U_0=0.1$ and $|V|/U_0=0.02$.
  In panel (a), the data are plotted near
  the tip of the $\rho=1$ Mott lobe of the phase diagram
  Fig.~\ref{fig:QMCPD}. AF order is destroyed at $t/U_0\sim 0.045$,
  {\it before} the system becomes a SF near $t/U_0\sim 0.06$.  The
  shaded region denotes the non-magnetic Mott insulator.  In panel
  (b), the same observables are plotted at $\rho=3$. Here the
  insulator-to-SF and magnetic transitions take place at the same
  critical values $t/U_0\sim 0.39$ and is first order.  }
\label{fig:SafSF.vs.t_rho1}
\end{figure}

To study magnetic properties of the model in QMC simulations, we
measure the real-space spin-spin correlation function along the
$z$-axis 
\be C_s(\bfr) = \frac{1}{L^2}\sum_{\bfr'} \ob{
  \hS^z_{\bfr+\bfr'}\hS^z_{\bfr'} }.  \label{eq:SpinCorr}
\ee 
Figure~\ref{fig:SpinCorr}
shows the results obtained for the $L=12$ lattice at $\rho=1$ (upper
panel) and $\rho=2$ (lower panel) with $U_2/U_0=0.1$, $|V|/U_0=0.02$, and
$t/U_0=0.02$, i.e. inside the MI phases. The staggered correlation
pattern displayed in Fig.~\ref{fig:SpinCorr} (a) shows that
the first Mott lobe is antiferromagnetic. Similar results are also
obtained for the third lobe.  As indicated by Fig.~\ref{fig:SpinCorr}
(b), the second Mott lobe at $|V|/U_0=0.02$ is non-magnetic since only
short-ranged correlation exists.

In order to confirm that the $\rho=1$ and 3 MIs have long-range
magnetic order, we have also studied the scaling of the spin structure
factor at $(\pi,\pi)$ \be S^{zz}(\pi,\pi)=\sum_\bfr (-1)^\bfr
C_s(\bfr).  \ee If the state has a long-range AF order, then
$S^{zz}(\pi,\pi)$ should scale as $L^2$\cite{Huse1988}. The results
for the first Mott lobe at $U_2/U_0=0.1$ and $|V|/U_0=0.02$ are plotted
in Fig.~\ref{fig:SafSF.vs.t_rho1} (a) as a function of $t/U_0$. In
this figure, the filled symbols represent $S^{zz}(\pi,\pi)/L^2$
computed for $L=6$, 8, 10, and 12. The data confirm that the first
Mott lobe has long-range AF order. By carrying out similar scaling
studies, we have verified that the third lobe at $U_2/U_0=0.1$,
$|V|/U_0=0.02$ (cf.  Fig.~\ref{fig:SafSF.vs.t_rho1} (b)), and the
$\rho=1$, 2, and 3 MI phases at $U_2/U_0=0.1$, $|V|/U_0=0.08$ also have
long-range AF order.  These findings confirm the MF predictions
regarding the magnetic structure of Mott insulators at commensurate
fillings in the parameter ranges studied.

 In addition to the AF order parameter $S^{zz}(\pi,\pi)$, we also
 show in Fig.~\ref{fig:SafSF.vs.t_rho1} the total SF density as a
 function of $t/U_0$ for $|V|/U_0=0.02$. It can be seen that at $\rho=1$, 
 the SF density $\rho_s$ rises and becomes
 size-independent (indicating a true SF phase) at $t/U_0\sim 0.06$, a value 
 that is consistent with the one found in Fig.~\ref{fig:QMCPD} (a).  
 Therefore, as one scans through $t/U_0$, the $\rho=1$ Mott insulator 
 undergoes a first order (indicated by the discontinuous jump in 
 $S^{zz}(\pi,\pi)/L^2$) magnetic phase transition at $t/U_0\sim 0.045$ 
 before it becomes a SF.
 This magnetic phase transition is not captured by the MF theory.

 Figure~\ref{fig:SafSF.vs.t_rho1}(b) shows a similar analysis near the
 tip of the third MI phase at $|V|/U_0=0.02$. It is found that the MI-SF
 transition is first order and takes place at $t/U_0\sim 0.037$. However,
 no intermediate phase exists.

\begin{figure}
\includegraphics[scale=0.4]{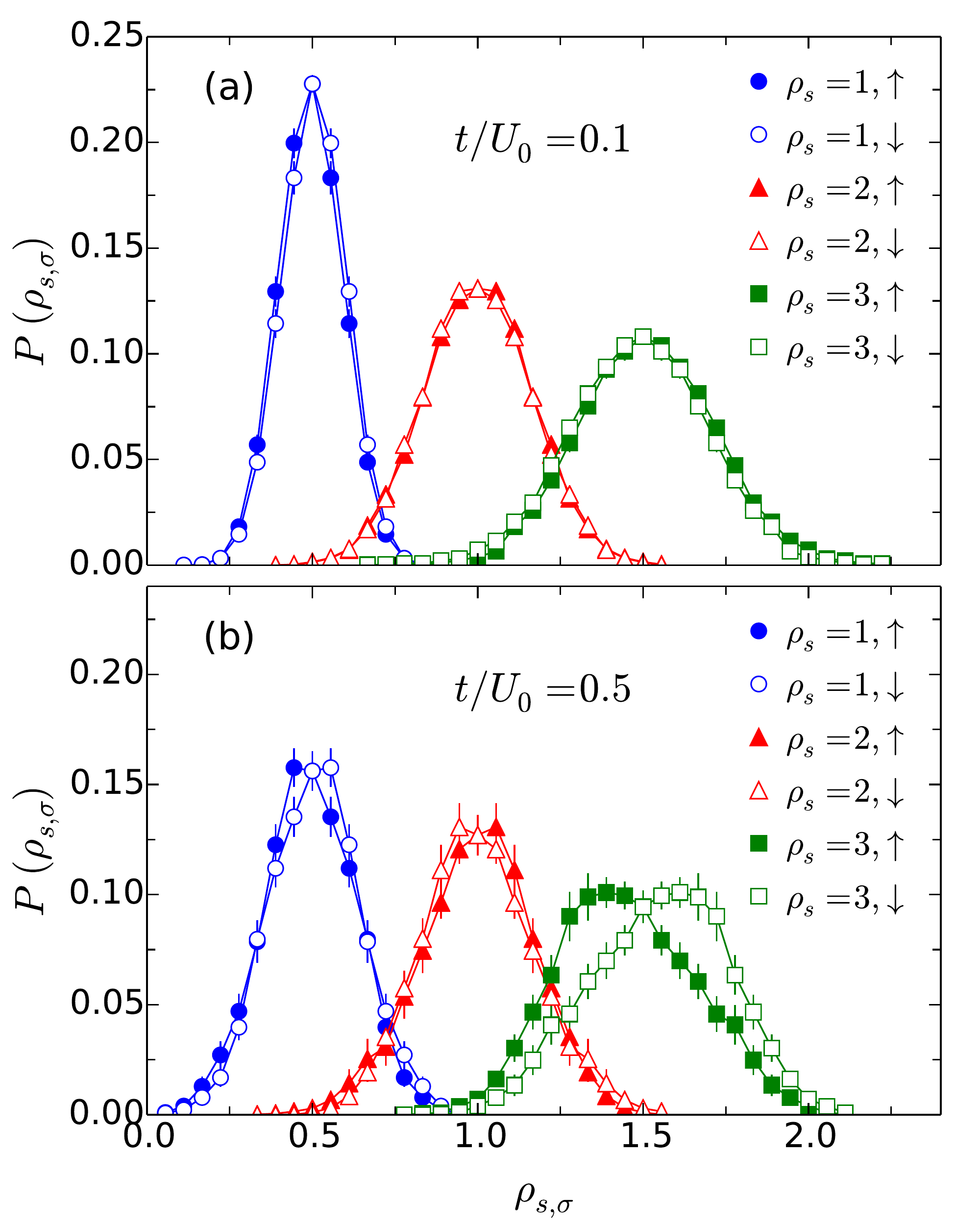}
\caption{(Color online) Histogram of the SF density
  $P(\rho_{s,\sigma})$ for
  $\sigma=\upa$ and $\dna$ bosons at three different
  commensurate fillings $\rho_s=\rho_{s,\upa}+\rho_{s,\dna}$.  The
  system is a $L=6$ lattice with $U_2/U_0=0.1$, $|V|/U_0=0.02$, and
  $\beta=12$.  In case (a) where $t/U_0=0.1$, the probability
  $P(\rho_{s,\sigma})$ centers at $\rho_s/2$, indicating that both
  spin components are equally populated in the SF state. In case (b)
  $t/U_0=0.5$, there is an asymmetry between $P(\rho_{s,\upa})$ and
  $P(\rho_{s,\dna})$ at $\rho_s=3$. This implies that the SF state has
  a finite polarization. }
\label{fig:Histo}
\end{figure}

\subsection{Magnetic properties of the SF phase}

Next we turn our attention to magnetic properties of the SF phase.
MFT predicts three different types of SF: a FMSF, an 
AFSF, and an unpolarized SF. At $|V|/U_0=0.02$, the FMSF dominates the
phase diagram. At a stronger NN spin coupling $|V|/U_0=0.08$, the AFSF
becomes the major component(cf. Fig.~\ref{fig:MFPD}).

To verify these MF predictions, we first compute the SF density histogram 
$P(\rho_{s,\sigma})$ for spin $\sigma=\upa$ and $\dna$ bosons.
As shown in previous results\cite{Larent:2010, Larent:2011}, $P(\rho_{s,\sigma})$ 
for $\sigma=\upa$ and $\dna$ are identical if both spin species are equally 
populated. On the other hand, $P(\rho_{s,\upa})$ and $P(\rho_{s,\dna})$ would 
peak at different values  of $\rho_{s,\sigma}$ if the superfluid develops 
polarization.

\begin{figure}
\includegraphics[scale=0.4]{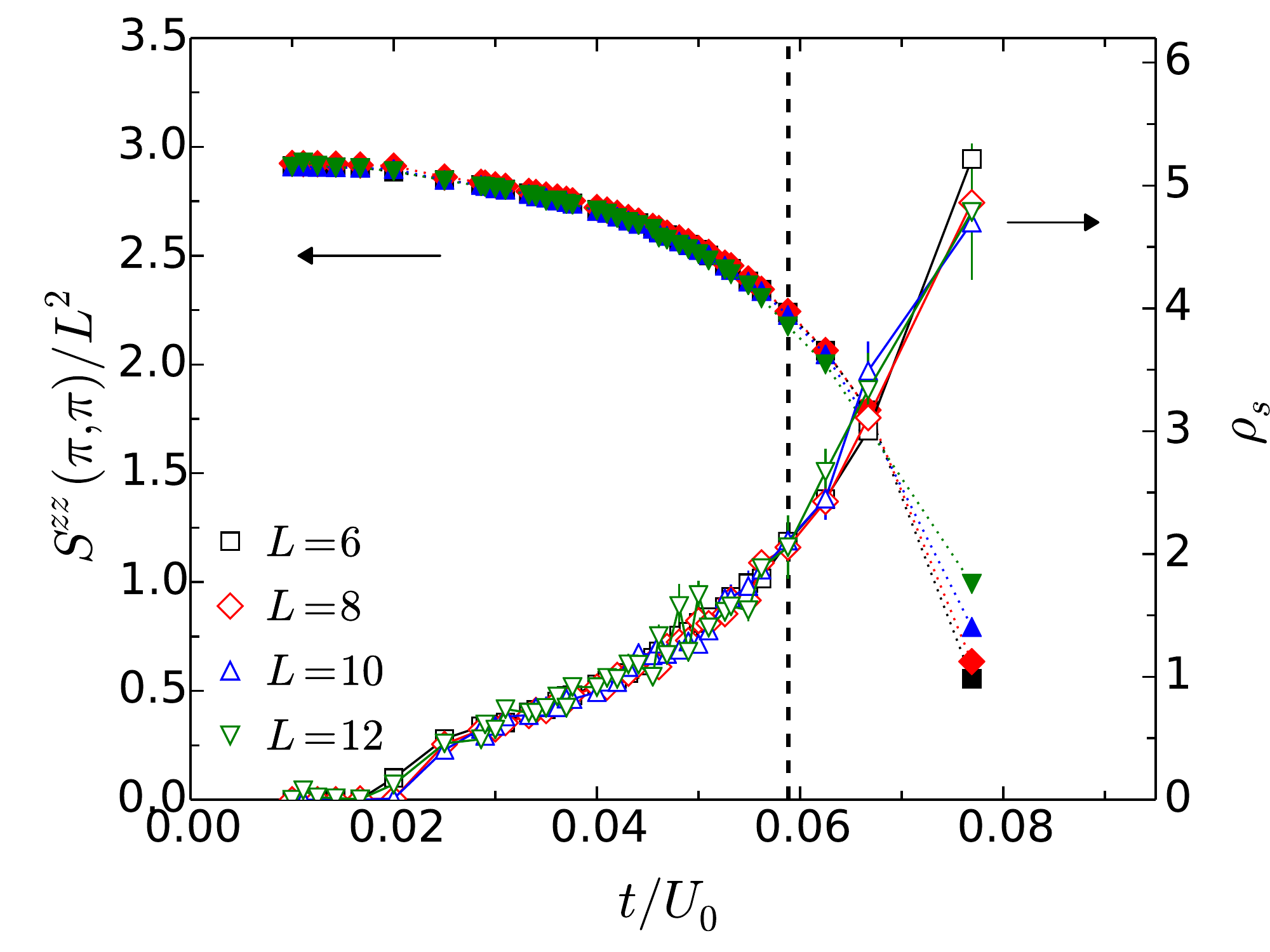}
\includegraphics[scale=0.4]{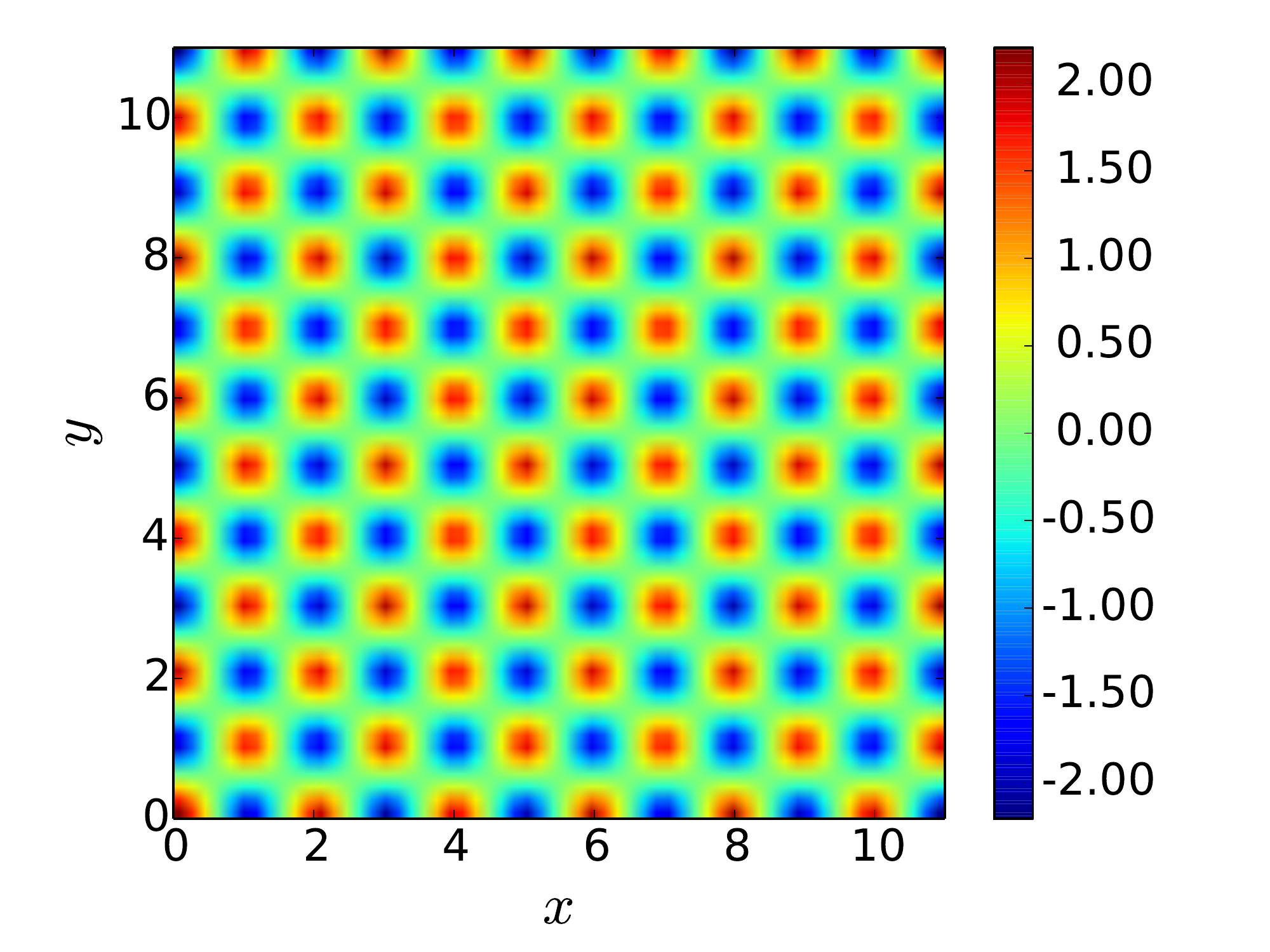}
\caption{(Color online)
Upper panel : Antiferromagnetic spin structure factor $S^{zz}(\pi,\pi)$
and SF density $\rho_s$  versus $t/U_0$ measured at $\rho=3.5$. The
onsite and NN spin interactions are $U_2/U_0=0.1$ and $|V|/U_0=0.08$
respectively.  The system is a SF for $t/U_0 \gtrsim 0.02$. Within the
SF phase, $S^{zz}(\pi,\pi)$ scales as $L^2$. These results suggest that
the SF has long-range AF order.  Lower panel: Spin-spin correlation
function $C_s(\bfr)$ computed on the $L=12$ lattice at the
$t/U_0=0.0588$, indicated by the dashed vertical line in the upper
panel.
}
\label{fig:SafSF.vs.t_rho3.5}
\end{figure}

At $U_2/U_0=0.1$ and $|V|/U_0=0.02$, examples of the histogram are
plotted in Fig.~\ref{fig:Histo}(a) for the $L=6$ lattice at three
commensurate densities $\rho_s = \rho_{s,\upa} + \rho_{s,\dna}=1$, 2,
3 at $t/U_0=0.1$, i.e. deep inside the SF phase. The figure shows that
$P(\rho_{s,\sigma})$ are identical for both spin components at a given
density and peaks at $\rho_s/2$, indicating no spin polarization.  We
find no FMSF in the parameter range shown in the $|V|/U_0=0.02$ QMC
phase diagram (top panel of Fig.~\ref{fig:QMCPD}).

In order to search for the FMSF further, we carry out the simulation
at much higher $t/U_0$ values.  One representative result of
$P(\rho_{s,\sigma})$ at $t/U_0=0.5$ is depicted in
Fig.~\ref{fig:Histo}(b) for the $L=6$ lattice with $U_2/U_0=0.1$ and
$|V|/U_0=0.02$.  The figure shows that at $\rho_s=1$, 2, and 3, the
histogram peaks at different locations for different spin
species. These results suggest the existence of a spin polarized SF
phase, albeit at a much higher hopping range than the MF prediction.
A similar conclusion is reached at $|V|/U_0=0.08$ for the FMSF phase.

\begin{figure}
\includegraphics[scale=0.42]{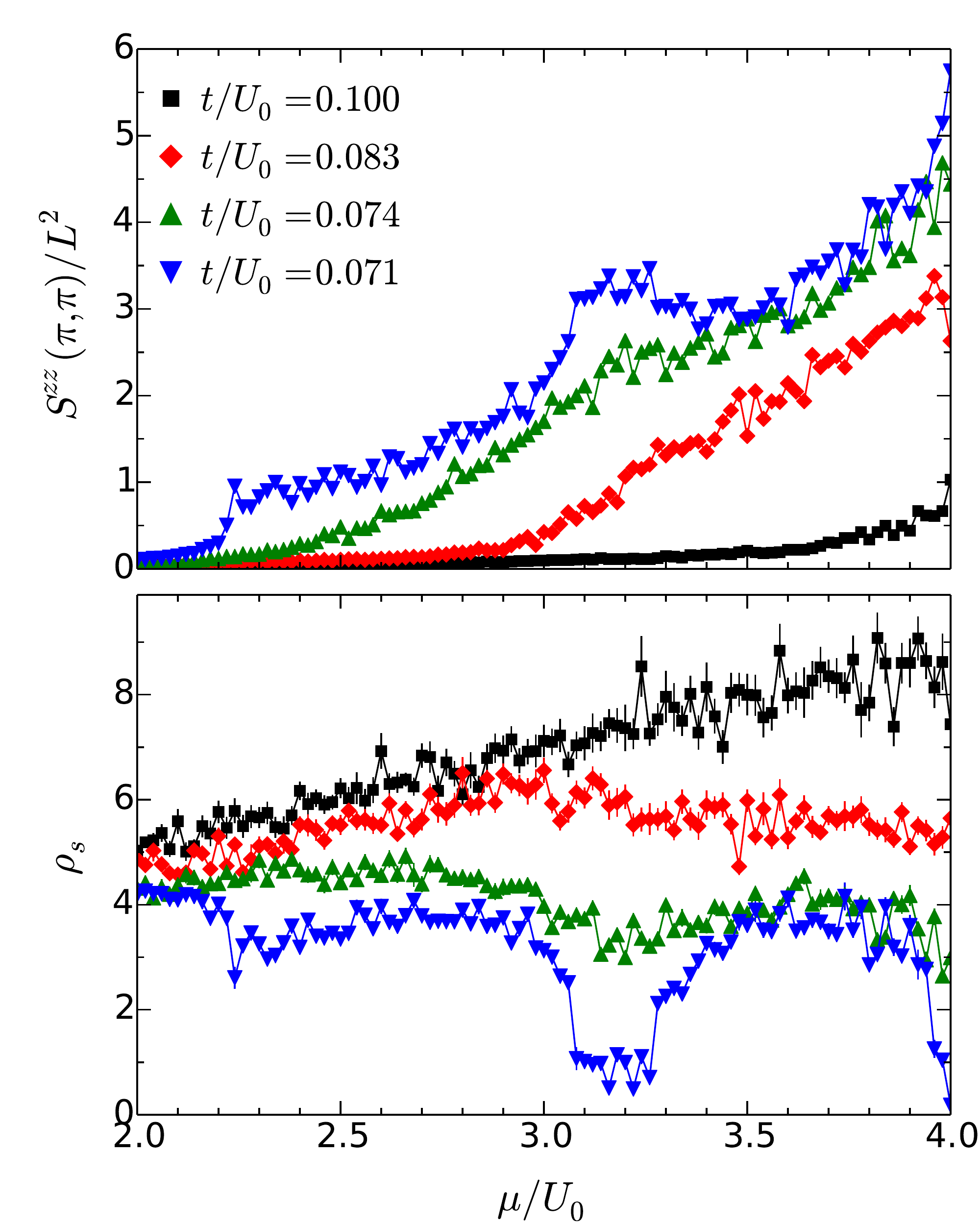}
\caption{(Color online)
Normalized spin structure factor at $(\pi,\pi)$ (top panel) and 
superfluid density (bottom panel) versus chemical potential $\mu/U_0$ 
computed on the $L=6$ lattice with $U_2/U_0=0.1$, $|V|/U_0=0.08$. 
In the top panel, there is a sudden increase in $S^{zz}(\pi,\pi)$ near
$\mu/U_0\sim 2.2$, 3.1, and 3.9 for $t/U_0=0.071$.
At roughly the same $\mu/U_0$ values and the same $t/U_0$, a reduction 
in $\rho_s$ can been observed.
}
\label{fig:Saf.vs.mu}
\end{figure}

To probe the AFSF phase, we compute the spin correlation function
Eq.~(\ref{eq:SpinCorr}) in the superfluid phase. At $|V|/U_0=0.02$,
results only indicate short-range AF correlations, and the
corresponding scaling study of AF spin structure factor does not
support any long-range AF order in the superfluid.

We carry out the same analysis for $|V|/U_0=0.08$, and the results at
$\rho = 3.5$ are summarized in Fig.~\ref{fig:SafSF.vs.t_rho3.5}.  The
upper panel of the figure shows $S^{zz}(\pi,\pi)/L^2$ as well as total
SF density $\rho_s$ in a range of $t/U_0$ values. The superfluid
density data indicate the onset of superfluidity is at $t/U_0 \sim
0.02$.
In the SF phase, the AF spin structure factor is finite and scales as
$L^2$ before it drops to zero at $t/U_0\sim 0.08$. These data combined
therefore confirm the existence of an AFSF phase at $|V|/U_0=0.08$.
This phase can be considered a supersolid phase since it exhibits
similtaneous digaonal and off-diagonal long range order. In the lower
panel of Fig.~\ref{fig:SafSF.vs.t_rho3.5}, we show a real-space spin
correlation function result acquired on an $L=12$ lattice at
$\rho=3.5$, $U_2/U_0=0.1$, $|V|/U_0=0.08$, and $t/U_0=0.0588$ (indicated
by the vertical dashed line in the upper panel of
Fig.~\ref{fig:SafSF.vs.t_rho3.5}). The staggered correlation function
pattern demonstrates the long range antiferromagnetic structure of the
SF phase. This long range AF order in the SF phase
  appeared as the NN repulsion was increased from $|V|/U_0=0.02$ to
  $|V|/U_0=0.08$. We have not, however, determined the value of $|V|/U_0$
  at which the AFSF first appears.

Figure~\ref{fig:SafSF.vs.t_rho3.5} also indicates that at some values
of $t/U_0$, the AF order vanishes and the SF becomes a normal
superfluid.  To estimate the exact phase boundary of this AFSF to
normal SF transition, we have conducted a series of grand canonical
and canonical SGF simulations and extracted $S^{zz}(\pi,\pi)$ as a
function of $\mu/U_0$ for several $t/U_0$ values.
A set of data is presented in Fig.~\ref{fig:Saf.vs.mu} for the $L=6$
lattice with $U_2/U_0=0.1$, $|V|/U_0=0.08$, and $2 \leq \mu/U_0 \leq 4$.
It can be seen from the figure that the transition from a normal SF to
AFSF takes place at a chemical potential value much higher than the MF
result, and the critical $\mu/U_0$ increases with $t/U_0$.  We have
done similar calculations for other system sizes.  
The estimated AFSF phase boundary is shown in Fig.~\ref{fig:QMCPD} (b).

Finally, we would like to remark that as one scans the chemical
potential in the range $2 \leq \mu/U_0 \leq 4$ at $t/U_0=0.071$, a
reduction in the SF density at $\mu/U_0\sim 2.2, 3.1$ and $3.9$ can
be observed in Fig.~\ref{fig:Saf.vs.mu}.  Correspondingly,
$S^{zz}(\pi,\pi)$ rises rapidly near these $\mu/U_0$ values.  Because
$t/U_0=0.071$ is just outside the tip of the third Mott lobe, the
reduction in $\rho_s$ at $\mu/U_0=2.2$ is caused by the proximity
effect of the third MI phase. The reduction in $\rho_s$ at
$\mu/U_0\sim 3.1$ indicates indirectly the location of the tip of the
fourth AF Mott lobe (and potentially the fifth at $\mu/U_0\sim 3.9$).

\section{Conclusion}  
  
In this work, we have used the site-decoupling MF theory and the exact
SGF QMC algorithm to study the ground state phase diagram of the
Bose-Hubbard model with onsite and NN spin-spin couplings
Eq.~(\ref{eq:BHspinHalf}).  The SGF approach allows us to treat terms
which interconvert the two bosonic species.  
Previous study\cite{Larent:2011}
have shown that the Hamiltonian at $V=0$ and positive $U_2$ has three 
phases: a ferromagnetic Mott insulator at $\rho=1$ (and all odd Mott 
lobes), an unpolarized Mott phase for $\rho=2$ (and all even commensurate
densities), and a ferromagnetic superfluid.

In the presence of NN interactions $|V|\sum_{\ob{\bfi\bfj}}\hat S^z_\bfi\hat S^z_\bfj$, 
the magnetic structures found at $V=0$ are profoundly modified.
In particular, at $|V|/U_0=0.02$, the Mott lobes at $\rho=1$, 3 
become antiferromagnetic. The Mott phase at $\rho=2$ is a spin-singlet
state. The superfluid phase becomes unpolarized for $t/U_0\lesssim 0.5$.
By increasing the strength of NN spin coupling to $|V|/U_0=0.08$,
the second Mott lobe also becomes antiferromagnetic, and, most interestingly,
an AFSF (a supersolid phase) emerges at high fillings.  

At the $|V|/U_0$ values
studied, the MF and exact QMC results are in good agreement,
particularly at small $t/U_0$ values (deep inside the Mott phase)
where quantum fluctuations are small.   
Moreover, the site-decoupling MFT is able to capture correctly the
magnetic structure of the Mott insulators and predict the existence of
AFSF. The order of MI-SF phase transition is also verified by the
exact results.

Just as initial qualitative studies of the single species
boson-Hubbard model were followed by quantitative comparisons with
experiment \cite{JimenezGarcia2010,Mahmud2011}, a natural next step
here will be to do similar modeling of multi-component bosonic optical
lattice experiments.  However, the complication introduced by the
effect of a trap, which in the single species case manifests itself as
the coexistence of superfluid, Mott insulator, and normal phases as
$\rho$, $U/t$ and $T/t$ vary across the cloud, will be even more
challenging, since the possibility of magnetic order introduces
additional phases which might coexist in the presence of a confining
potential.


\begin{acknowledgements}
The authors are grateful for support from the University of
Nice -- U. C. Davis ECOPAL LIA joint research grant, NSF-PIF-1005503 and
DOE SSAAP DE-NA0001842.
\end{acknowledgements}

\bibliography{reference}

\end{document}